\documentclass[aps,prb,preprint,showpacs,groupedaddress]{revtex4-1}
\usepackage{graphicx}% Include figure files
\begin{document}

%\preprint{}

\title{Octagonal Family of Monolayers, Bulk and Nanotubes}

\author{Prashant Vijay Gaikwad}
%\affiliation{Department of Physics and Centre for Modeling and Simulation, Savitribai Phule Pune University, Pune 411007, India}
\email{prashant@physics.unipune.ac.in}
\author{ Anjali  Kshirsagar}
\affiliation{Department of Physics and Centre for Modeling and Simulation, Savitribai Phule Pune University, Pune 411007, India}
\email{anjali@physics.unipune.ac.in}

\date{\today}

\begin{abstract}
A new class of tetragonally symmetric 2D octagonal family of monolayers ({\it{o}}-MLs) has emerged recently and demands understanding at the fundamental level.
{\it{o}}-MLs of metal nitride and carbide family (BN, AlN, GaN, GeC, SiC) along with C and BP are computationally designed and their stability and electronic 
structure are investigated.
These binary {\it{o}}-MLs show mixed ionic and covalent bonding with the hybridized {\it{p}} states dominating the electronic structure around the Fermi level.
Geometric and structural similarity of {\it{o}}-C and {\it{o}}-BN has been exploited to form patterned hybrid {\it{o}}-MLs ranging from metallic to insulating phases.
Stacking of zigzag buckled {\it{o}}-MLs results in stable body centered tetragonal (bct)-bulk phase that is suitable for most materials from group IV, III-V and II-VI.
Vertically cut chunks of {\it{o}}-BN and {\it{o}}-C bulk or stacking of {\it{o}}-rings, unlike rolling of hexagonal ({\it{h}})-ML, provide a plausible way to 
form very thin {\it{o}}-nanotubes ({\it{o}}-NT).
Confined and bulk structures formed with an octagonal motif are of fundamental importance to understand the underlying science and for technological applications.
\end{abstract}

%\pacs{31.15.ar, 61.46Fg, 71.20Nr, 73.22.-f, 73.63.Fg}
%\keywords{Octagonal monolyers, bct bulk, BN nanotubes, density functional theory}
\maketitle

\section{\label{sec:level1}Introduction}
Two-dimensional (2D) monolayers (MLs) are at extreme technological limits for next generation applications. Flat and planar structures
have properties and functionalities that can exploit new avenues for research and technological applications.
With the technological advent, tremendous interest has emerged in their synthesis and in engineering their properties for applications in electronics, energy
storage, high mechanical strength materials and so on.
These 2D MLs and their patterned hybrids and multilayered assemblies (graphitic BN, transition metal monochalcogenides and dichalcogenides, silicene, phosphorene) have
also gained importance for advanced strain driven technologies (straintronics and twistronics) because of their inherent crystallographic
and electronic structural differences.~\cite{Novoselov, Meunier, Pereira, Naumis}
Such 2D materials are reported to show versatile physical phenomena ranging from quantum Hall effect fractal spectrum, reversal of Hall effect,
external dielectric environment driven fluctuations in local Coulomb interaction, anomalous spin-orbit torque, etc. on one hand,
while applications in 2D electronic devices, spin-polarization switches via strain, giant pseudo-magnetic fields, giant 2D band-piezoelectric effect by biaxial-strain, 
and so forth on the other.~\cite{Dean, Liu2, Diniz, Ong, Levy, Xiaomu, Raja, Wang}
This demands advancement towards fundamental understanding of existing materials and exploring new possibilities of 2D materials having different structural forms.
Predictions of new phases of 2D materials can provide impetus for obtaining exotic properties and possibilities to overcome present synthesis challenges.

Various approaches have been developed to predict new possible allotropes of confined structures based on evolutionary studies,
however most of them require considerable computation.~\cite{De}
We have reported octagonal ZnO ({\it{o}}-ZnO) ML through high coordinated cluster assembly route as the next stable plane after the most stable hexagonal 
ML ({\it{h}}-ML).~\cite{Prashant}
This higher coordination fundamental approach provides use of passivating ligands mimicking the bulk-like environment, and demands comparatively less expensive
computational process using the ground state geometries of small-sized bare clusters (X$_{n}$ or A$_{n}$B$_{n}$ for $n$ atom elemental cluster of species 
X or $n$ atom stoichiometric
cluster of compound AB).
These cluster geometries can be used as building blocks to predict new stable geometries of 2D materials.
The geometries of 2D allotropes using pentagons and hexagons as unit building blocks are known and have been studied for various electronic and mechanical 
properties.~\cite{Sahin, Zhang}
Pentagons in such works are inspired by the synthesis of pure pentagon based C$_{20}$ cage and Cairo pentagonal tiling resulting in T12-carbon.
The possibility of C in octagonal ML ({\it{o}}-ML) form has also been reported,~\cite{Liu,Sheng} however the geometry has been chosen differently.

It is known that group IV, III-V and II-VI materials adopt either cubic zinc blende or hexagonal wurtzite crystallographic structures in three
dimensions. Many of these have been reported to form hexagonal two-dimensional monolayers. We have formed, for scientifically important group of 
materials, {\it{o}}-MLs using cluster assembly route.
Out of a large pool of investigated materials, ZnO, C, BN, SiC, GeC, AlN, GaN and BP have shown the possibility to form stable {\it{o}}-MLs.
Most of the {\it{o}}-MLs of various materials, proposed in the present study, are known to have hexagonal 2D allotropes as their ground state.~\cite{Topsakal, Sahin}
Our preliminary calculations for the {\it{o}}-MLs of BAs, AlP, MoS, CdO, CdS, CdTe, CN, CuO, GaAs, GaP, InN, CdTe, LaN and LaO have not shown dynamical stability.
Some recent reports have suggested that {\it{o}}-MLs of N, P, B, As, Sb, BAs and AlP are possible stable structures.~\cite{Zhang2, Brown}
however our phonon dispersion calculations suggest that those are dynamically unstable planes.
To further investigate fundamental aspect of evolution, we have formed bulk phases by stacking respective stable {\it{o}}-MLs.
This body centered tetragonal (bct)-bulk phase is found to be stable for all proposed {\it{o}}-MLs.
The most stable and experimentally easy to cleave {\it{o}}-MLs of C and BN are further exploited to form nanotubes and they appeared to be dynamically stable 
and show metallic and semiconducting nature respectively.

\section{Computational details}
Density functional theory (DFT)~\cite{Hohenberg, Kohn} based first principles calculations are performed using Vienna ab-initio simulation package 
(VASP).~\cite{Kresse1,Kresse3,Kresse4,Kresse}
Plane augmented wave (PAW) method~\cite{Blochl} as implemented in VASP is used with exchange-correlation energy functional proposed
by Perdew, Burke and Ernzerhof (PBE) within the generalized gradient approximation (GGA).~\cite{Perdew,Perdew1}
A 15$\times$15$\times$1 mesh of $\vec{k}$-points is used for primitive cell (8~atoms) of $\textit{h-}$ and {\it{o}}-MLs while 7$\times$7$\times$1 
and 5$\times$5$\times$1 sized meshes are used for 2$\times$2$\times$1 (32~atoms) and 3$\times$3$\times$1 (72~atoms) supercells of {\it{o}}-MLs of 
BN containing defects and/or dopants respectively.
For the bulk structures and nanotubes, 15$\times$15$\times$15 and 1$\times$1$\times$9 $\vec{k}$-point meshes are used respectively.
In all calculations vacuum of at least 18~\AA\ is maintained along the $\textit{z}$-direction for planar structures (in the $\textit{xy}$-planes) 
while along $\textit{x}$- and $\textit{y}$-directions for nanotubes (oriented along $\textit{z}$-direction).
The electronic energy and ionic forces are converged up to 10$^{-4}$~eV and 10$^{-3}$~eV/\AA\  respectively for all calculations.

The vibrational spectra calculated at $\Gamma$ and phonon dispersion spectra of respective unit cells, using Phonopy code~\cite{Phonopy}, are used 
as a first filter to explore the stability of {\it{o}}-ML structures.
After that a 3$~\times$~3~$\times$~1 supercell is used for phonon dispersion calculations for planes and 1$\times$1$\times$3 supercell is used for 
nanotubes to confirm the dynamical stability and to calculate thermal properties and phonon partial density of states.

Cohesive energy (E$^{c}$) of a given ML is calculated using Eq.~(\ref{e1}), where E$_{X_{n}Y_{n}}$ is the total energy of {\it{o}}-XY with $n$
atoms of X and Y species.
E$_{X}$ and E$_{Y}$ are energies of isolated X and Y type of atoms respectively.
E$^{c}$/$n$ is the cohesive energy per pair (E$^{c}$/XY) and is twice E$^{c}$/atom.
Positive value of E$^{c}$ indicates stable {\it o}-ML structure formation.

Substitutional energy (E$_{sub}$) to substitute $q$ atoms of species Y by species T is calculated using Eq.~(\ref{e2}).
The term E$_{X_{n}Y_{n-q}T_{q}}$ refers to the energy of the structure with substitution of $q$ atoms of species T for species Y, whereas 
E$_{X_{n}Y_{n}}$ is the energy of pristine structure.
More positive values of E$_{sub}$ represent comparatively higher stability of the substituted structure.
Cleavage energy with respect to respective {\it{o}}-bulk phase is defined by Eq.~(\ref{e3}), where the terms E$^{o-ML}$ and 
E$^{o-bulk}$ refer to the respective total energy per XY pair of {\it{o}}-ML and {\it{o}}-bulk phase.
The {\it{o}}-bulk phase is the optimized structure obtained after stacking {\it{o}}-MLs.

\begin{eqnarray}
\label{e1}
E^{c} &= &n * E_{X} + n * E_{Y} - E_{X_{n}Y_{n}}\\
\label{e2}
E_{sub} &= & [E_{X_{n}Y_{n}} - q * E_{Y} + q * E_{T})] - E_{X_{n}Y_{n-q}T_{q}}\\
\label{e3}
E^{clv} &= & E^{{\it{o}}-ML} - E^{{\it{o}}-bulk}
\end{eqnarray}

\section{Results and Discussion}
\begin{figure}[!ht]
\centering
\includegraphics[scale=.82]{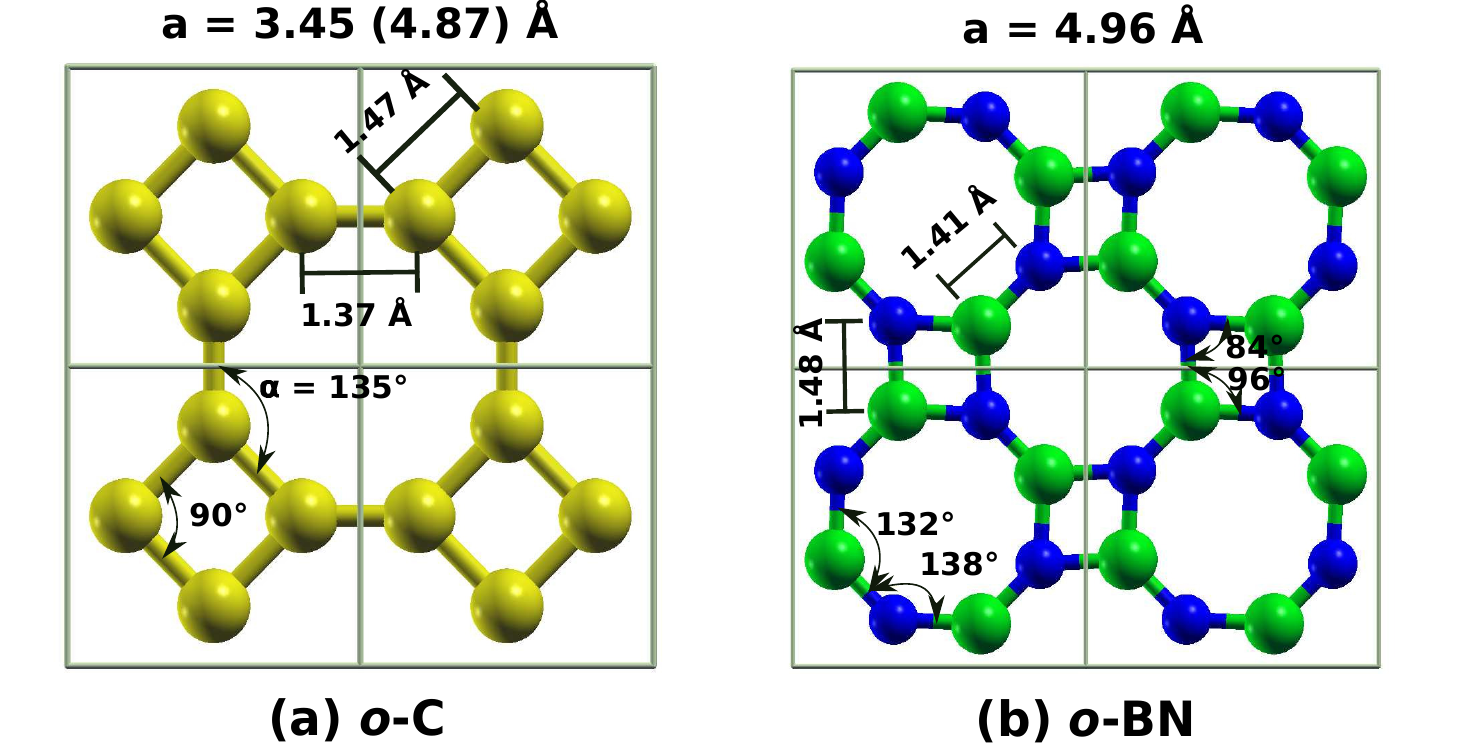}
\caption{A 2~$\times$~2~$\times$~1 supercell of {\it{o}}-MLs of (a)~C and (b)~BN. The lattice parameter {\it{a}} of {\it{o}}-ML of C
for larger {\it{o}}-BN like unit cell is also shown in parenthesis for comparison. Representative bond lengths and angles are shown. Yellow, blue, 
and green balls represent C, N and B atoms respectively.}
\label{geo1}
\end{figure}

2~$\times$~2~$\times$~1 supercells of {\it{o}}-MLs of C and BN are shown in Fig.~\ref{geo1}.
Similar to {\it{h}}-ML structure, the {\it{o}}-ML structure is a 3 coordinated planar structure.
{\it o}-BN can be viewed as the structure formed out of combination of passivated cluster geometries 
obtained from ground state geometries of B$_2$N$_2$ and B$_4$N$_4$ clusters.
All the reported  {\it{o}}-MLs in this work are designed using cluster assembly route as reported in our earlier work.~\cite{Prashant}
The planar structure of {\it{o}}-ML of C belongs to P4/{\it{mmm}} (123) while that of BN belongs to P4/{\it{mbm}} (127) tetragonal space group.
As evident from Figs.~\ref{geo1} (a) and (b), {\it{o}}-ML of C is a more symmetric structure. {\it{o}}-ML structure can be viewed as the 
structure formed out of planar arrangement of octagonal rings (passivated cluster geometry of C$_{8}$ separated by 1.47~\AA\ in case of {\it o}-C 
and that of B$_{4}$N$_{4}$ separated by 1.48~\AA\ in case of {\it o}-BN).
The bond angles inside the octagonal region are not exactly 135$^{\circ}$ and those inside the square region are not exactly 90$^{\circ}$ for {\it{o}}-BN, as is the
case for {\it o}-C.
These facts generate asymmetry in the crystal structures of various binary compounds and restrict the choice of elements.

\begin{table}[!ht]
\caption{Variation of E$^{c}$ per pair (E$^{c}$/XY in eV), cleavage energy per pair (E$^{clv}$/XY in eV), lattice parameter ({\it{a}} in \AA), 
bond lengths ({\it{d}} in \AA) and bond angles ($\alpha$ in degree) of octagon in respective {\it o}-MLs.}
\begin{center}
\begin{tabular}{c c c c c c c}
\hline \hline
System & E$^{c}$/XY & E$^{clv}$/XY & E$_{g}$ & {\it{a}} & {\it{d}} & $\alpha$ \\
 & (eV) & (eV) & (eV) & (\AA) & (\AA) & (in degree) \\
\hline
Graphene & 15.98 & 0.02\footnote{ The cleavage energy of graphene is with respect to the graphite structure} & 0.00 & 2.47 & 1.43 & 120.0 \\
C & 14.95 & 0.14 & 0.00 & 4.87\footnote{ The lattice parameter of {\it{o}}-ML of C is defined in terms of 8 atom tetragonal unit cell while the 
primitive cell lattice parameter is 3.45~\AA} &  1.37, 1.47 & 135.0 \\
SiC & 11.48 & 1.16 & 4.06 & 6.09 & 1.75, 1.81 & 132.8-137.2  \\
GeC & 9.32 & 1.08 & 1.72 & 6.45 & 1.84, 1.92 & 134.8-135.2  \\
BN & 13.57 & 0.08 & 4.13 & 4.96 & 1.41, 1.48 & 132.2-137.8  \\
AlN & 10.16 & 1.11 & 2.86 & 6.16 & 1.76, 1.83 & 133.1-136.9  \\
GaN & 7.64 & 0.93 & 1.82 & 6.42 & 1.83, 1.91 & 134.9-135.1  \\
BP & 9.41 & 0.98 & 0.79 &  6.35 & 1.82, 1.89 & 134.8-135.2  \\[1ex]
\hline
\end{tabular}
\end{center}
\label{stability}
\end{table}

\begin{figure}[!hb]
\centering
\includegraphics[scale=.62]{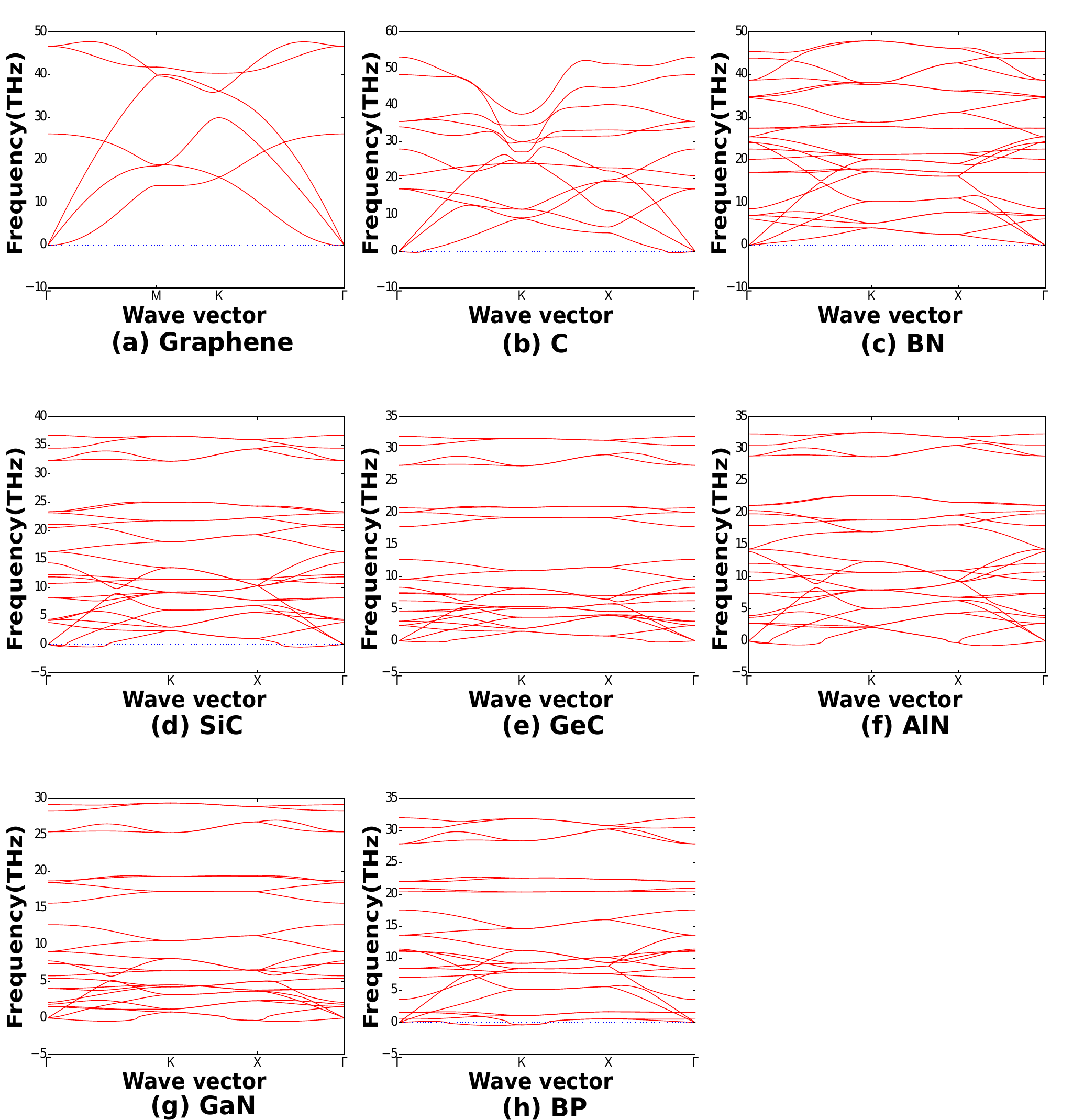}
\caption{Phonon dispersion spectra of (a)~graphene and ((b) through (h))~{\it{o}}-MLs of C, BN, SiC, GeC, AlN, GaN and BP respectively.}
\label{phon-stab}
\end{figure}

The stability of these structures is studied by investigating structural, energetic and dynamical stability.
Table~\ref{stability} lists the E$^{c}$ per pair, optimized lattice parameter (refer to SI for typical graph for {\it o}-GaN), bond lengths and bond 
angles of respective stable {\it{o}}-MLs. Here in the table and at other places in this report, results for graphene have been included for 
comparison. These results agree with the corresponding results published earlier in the literature.
%values of E$^{c}$ per pair in case of {\it{o}}-ML of C and graphene are also listed.
>From the table it is clear that among all planes, {\it{o}}-ML of BN is the next stable structure after C and it has bond lengths close to graphene.
Among each family of carbide and nitride, stability decreases and bond length increases with increasing Z.
{\it{o}}-ML of GaN has least E$^{c}$ although it has comparable Z and lattice parameter with GeC. This can be attributed to asymmetric {\it{sp$^{2}$}} 
hybridization in GaN.
Table~\ref{stability} lists the respective cleavage energy per pair (E$^{clv}$/XY) of various {\it{o}}-MLs from their respective {\it{o}}-bulk structures.
Lower cleavage energy of 0.14~eV and 0.08~eV for {\it{o}}-MLs of C and BN from respective bulk structures suggests strong possibility of experimental cleavage.

The calculated phonon dispersion spectrum for graphene is shown in Fig.~\ref{phon-stab} (a).
The lower frequency modes are acoustic (ZA, TA, LA) modes while higher frequency modes are optical (ZO, TO, LO) modes.
The calculated degenerate LO and TO modes of graphene at $\Gamma$ are around 1600 cm$^{-1}$, which are in agreement with previously reported theoretical 
and experimental values.~\cite{Sahin,Lee,Narasimhan,Yan}
The phonon dispersion spectra of {\it{o}}-MLs of C, BN, SiC, GeC, AlN, GaN and BP are shown in Figs.~\ref{phon-stab} ((b)-(h)),
while respective partial density of states ({PDOS) contributed by the constituent atoms can be found in SI.
To include large wavelength modes of phonons, a 3~$\times$~3~$\times$~1 unit cell is chosen in all cases.
It shows minor soft phonon modes in some cases.
The imaginary frequencies around $\Gamma$ point and their extent for acoustic modes can be an artifact of limited accuracy of the calculations.
It is known that for 2D confined systems the bending branch ZA can easily become soft and hence may show imaginary frequencies, this however 
contributes negligibly to the thermal conductivity.~\cite{ Klemens} The soft phonon modes causing instability in the {\it o}-ML structures
reflect the stress induced in the {\it o}-ML. These may be related to significantly different hybridization of deep lying electronic states as well as
the electronic states lying near the Fermi energy, in comparison to the stabilization mechanism in respective bulk structures, as will be discussed
in subsection~\ref{EleSt}. As a result, 
such {\it{o}}-MLs show significantly different properties.
Structures, which show soft phonons, are expected to be more stable in multilayer form.
It is indeed reported earlier for {\it{o}}-ZnO.~\cite{Prashant}

All the {\it{o}}-MLs of binary materials belong to tetragonal space group P4/{\it{mbm}} (127) and symmetry of the structure demands crossing of ZA 
and OA modes at symmetry points K and X resulting in degenerate states along K~-~X path.
Among the {\it{o}}-MLs studied, C and BN are the most stable. {\it{o}}-MLs of SiC and GeC are dynamically stable and those of AlN and GaN show soft phonons.
The phonon dispersion spectra for unit cell of {\it{o}}-ML for various materials are provided in supporting information.
It is seen that {\it{o}}-MLs of BN, AlN, GaN, BP, GeC, SiC, C are stable.
However, for long wavelengths, {\it{o}}-MLs of BP and GaN show soft phonons (Figs.~\ref{phon-stab} (g) and (h)).
Recently, {\it{o}}-ML of GaN and its bulk structure are reported to show no soft phonons (calculated using SIESTA code).
The reported optimized bond lengths are marginally greater than those found in the present work but the cohesive energy
per atom found in present work suggests a structure more stable than reported earlier.~\cite{Mojica}
In the present work, initial guess for geometry is chosen based on trial run of respective passivated clusters.
Initial guess is critical since optimization process is complex and difficult due to shallow minimas in the potential energy surface.
The phonon dispersion spectrum of {\it{o}}-ML of GaN primitive cell (refer SI) is qualitatively in agreement with the previous reports~\cite{Mojica},
however calculations of phonon spectra using 3~$\times$~3~$\times$~1 supercell shows soft phonons governed by ZA modes as shown in Fig.~\ref{phon-stab} (g).
The phonon PDOS plots of {\it{o}}-MLs of C, SiC and GeC (refer to SI)
clearly show that the induced softening of ZA is caused by heavier atoms and it increases with atomic number. Similar results can be seen for nitride family (refer SI).
Since the force constant increases with atomic number, the lighter elements contribute dominantly to the high frequency modes.
The structures which are not stable at longer wavelength phonon modes can become more stable if defects are introduced to restrict the propagation of modes.

It is to be noted that structures containing the lighter first row elements (B, N, O) prefer to form more stable planar {\it{o}}-MLs.
Similar observations are reported for hexagonal structures and other carbon allotropes.~\cite{Sahin}
The elements having {\it{s}} and {\it{p}} valence states can comparatively easily form planar bonds and therefore form stable {\it{o}}-MLs.
Non-planar hybrid states of {\it{p}}, {\it{d}} or {\it{f}} may give rise to dangling bonds.
One way to overcome this problem is by inducing buckling in the structure, which allows such $\textit{z}$-directional states
to form bonds and hence increase the stability of the structure.
However, in {\it{o}}-ML structure, the geometrical flexibility of such buckling is restricted to few choices.
One possibility corresponds to buckling of alternate pairs of atoms along {\it{z}}-axis in opposite directions by equal amount leading to
bulk T-carbon or bct-C$_{4}$ structure.~\cite{Umemoto}
Another possibility is zigzag buckling of alternate atoms of octagon (4 up and 4 down), known as bct-C$_{8}$ structure.
It amounts to stacking of two planes which are mirror images of each other to form bilayer. It may lead to a bulk phase as discussed in subsection~\ref{bulk1}.
The same bulk phase can be obtained by stacking nanotubes as explained in subsection~\ref{ntube}.

Both, planar and buckled {\it{o}}-Ge and {\it{o}}-Si are energetically stable but are dynamically unstable.
Surprisingly, these octagonal planar structures, though consisting of single type of atoms from group IV, show inherent ionicity.

With increase in the atomic number the force constant decreases, which is compensated by stronger bonding and decrease in the bond length.
This is confirmed by the observed variation in the phonon DOS with atomic number.
However, with increase in atomic number the ionic radius increases which demands higher lattice parameter for stability, particularly at the square part of {\it{o}}-MLs.
Since these criteria cannot be fulfilled by transition metals (higher Z, presence of {\it{z}}-directional states and higher ionic radius) they do not form stable {\it{o}}-MLs.
However, we argue that two or more layered assemblies of such materials can lead to displacive phase transition led stable structures.
Indeed this is observed in our previous reported work on {\it{o}}-ZnO.~\cite{Prashant} The bulk form T-carbon or Haeckelite structures can be useful in this direction.

Thermal properties \textit{viz}. Helmholtz free energy, entropy and specific heat of various {\it{o}}-MLs are calculated and are 
incorporated in supporting information.
It is evident that the crossover of specific heat with entropy shifts to higher temperature with the stability of the structure as reflected by respective E$^{c}$.
It may be noted that with increase in atomic number, the crossover temperature decreases.
Thus the stability inferred from E$^{c}$, phonon dispersion plots and atomic numbers of constituent materials agree with the decrease in the crossover temperature.

\subsection{Electronic structure of {\it{o}}-MLs}\label {EleSt}

Figure~\ref{band} depicts the electronic band structure of {\it{o}}-MLs.
The electronic structures of graphene and {\it{o}}-ML of C and BN agree with respective previous reports.~\cite{Liu, Brown, Shahrokhi}
The band structures show that states around Fermi energy are principally constituted by the {\it{p}} states, in particular the $p_z$ 
states, of the constituent elements. The strength of the bonding and antibonding, $\pi$ and $\pi^*$, states depends on the distance
between the parent atoms. If one of the element in the structure is from the first row, size of the atom is small and therefore strong
$\pi$ bonds exist in the system. In ionic materials, the $p_z$ states mostly lie on the same atom, the degeneracy of states is lifted
and the valence and conduction bands separate giving finite band gap. Charge transfer and hence ionicity decreases with increasing Z.
Thus hybridization of $p$ states plays a dominant role and changes with Z and size of atom resulting in different electronic properties.

\begin{figure}[!h]
\centering
\includegraphics[scale=.62]{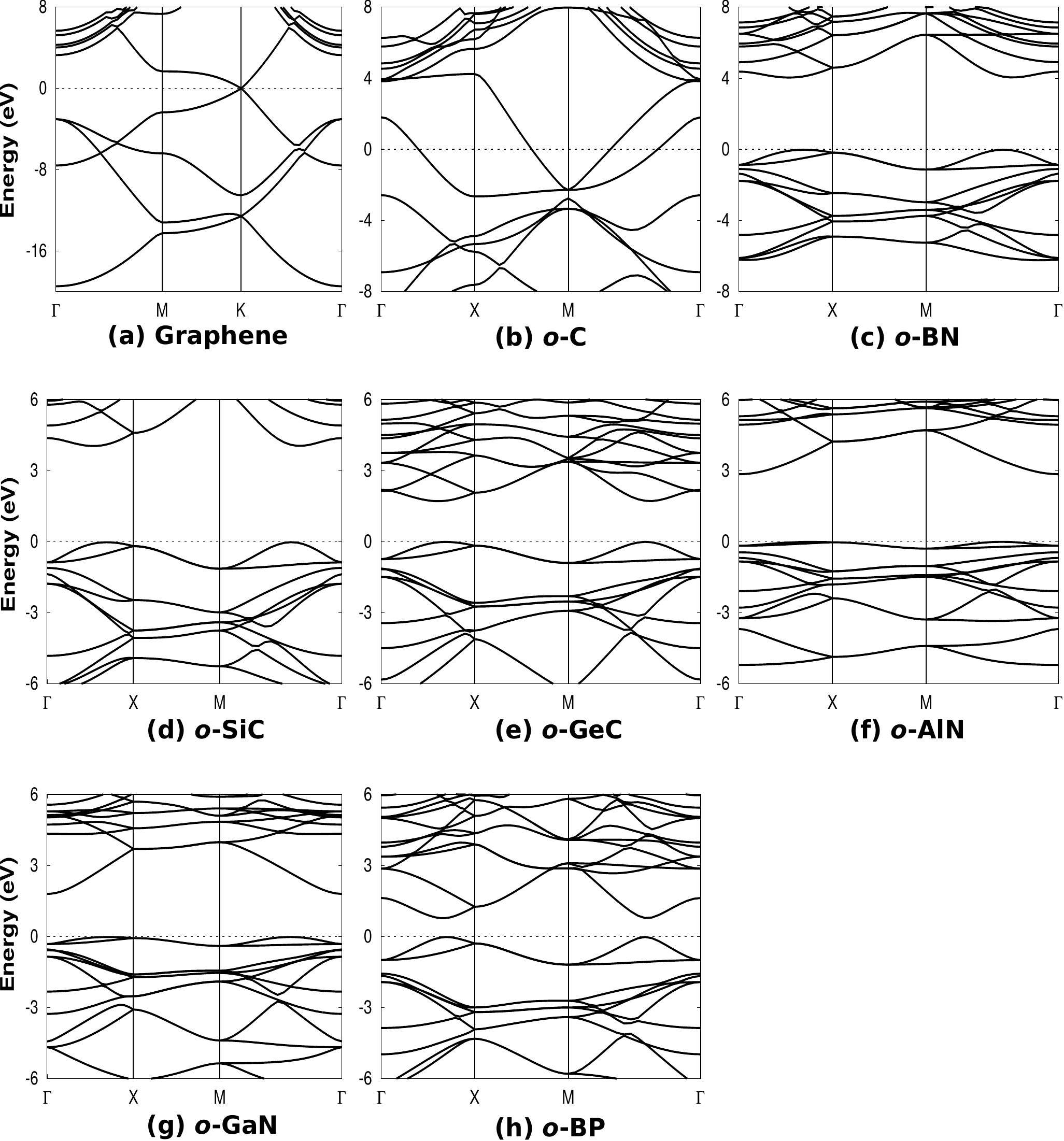}
\caption{Electronic band structures of (a)~graphene and ((b)-(h)) {\it{o}}-MLs of C, BN, SiC, GeC, AlN, GaN and BP.}
\label{band}
\end{figure}

The Dirac cone of graphene is shifted towards lower energy for {\it{o}}-C and this results in metallic nature.
All other structures show a finite band gap in the range of 4 to 0.8~eV as shown in Table~\ref{stability}.
BN, SiC, GeC and BP show direct band gap while AlN and GaN show quasi-direct band gap.
GeC, AlN and GaN have band gap in the visible range and hence can be promising candidates for photo-electronic applications.
Furthermore, with introduction of impurities, defects or doping and stacking of layers, the band gap is likely to decrease and can be tuned; 
in particular the large band gap ($\sim$4~eV) of BN and SiC is interesting.
{\it{o}}-ML of C is energetically the most stable and shows large number of conducting states at Fermi energy in comparison to graphene (Figs.~\ref{band} (a) and (b)).
This makes {\it{o}}-ML of C a preferable candidate for applications wherein higher conductivity is required.
On the other hand, large stability (evident from E$^{c}$, phonon dispersion spectrum and thermal properties) and in-built band gap makes {\it{o}}-ML of BN
an excellent candidate for applications wherein large band gap is required.
Moreover, considering the nearly equal binding energy and lattice parameter of {\it{o}}-ML of C and BN, the band gap of {\it{o}}-ML of BN can be further tuned by
C doping and is discussed in subsection~\ref{hybrid}.

Charge density plots and Bader charge analysis of {\it{o}}-MLs confirm strong covalent nature of {\it{o}}-C and ionic nature of {\it{o}}-BN.
{\it{o}}-BN is ionic because B, with low electronegativity, prefers to donate three valence
electrons (Q$^{B}$ = +3$e$) while N accepts average charge (Q$^{N}$) of -3{\it{e}}.
The N sites of octagonal geometry are less symmetric compared to graphene-like hexagonal structure and have
either -2.89{\it{e}} or -3.11{\it{e}} charge on alternate near neighbor N atoms.
We ascribe this asymmetry to asymmetric bonding of {\it{p}} states in the {\it{xy}}-plane. 
As expected {\it{o}}-C is covalently bonded but there is a small charge transfer on alternate near neighbors.
This may likewise be due to asymmetric {\it{sp$^{2}$}} bonding (refer to SI for details).
However, mention may be made that each C$_{8}$ ring is overall neutral. The actual value of charge transfer may be in error 
as is well-known for graphene.~\cite{web} 
We have seen similar charge transfer in graphene and {\it o}-C but {\it o}-C being less
symmetric, we expect some ionicity in bonding.
To verify further, we have investigated ionicity of dynamically unstable {\it{o}}-Si and {\it{o}}-Ge. As expected, both {\it{o}}-MLs show 
partial ionicity in bonding.
Similar to {\it{o}}-C, Ge shows small charge transfer on alternate atoms and is covalently bonded, while {\it{o}}-Si is highly ionic with charge 
transfer as high as $\pm$1.3{\it{e}}.
Although {\it{o}}-Si is dynamically unstable, the calculated ionicity, as high as ZnO, makes it important for hydrogen storage and water 
splitting applications and needs attention.
Camacho-Mojica {\it et al.} have shown in their work on hexagonal and Haeckelite GaN that the charge 
transfer between Ga and N is not much different in the two cases.~\cite{Mojica} We therefore expect the same to hold for the {\it o}-MLs studied in this work.

\subsection{Hybrid {\it{o}}-MLs of BNC}\label {hybrid}

\begin{figure}[!h]
\centering
\includegraphics[scale=.68]{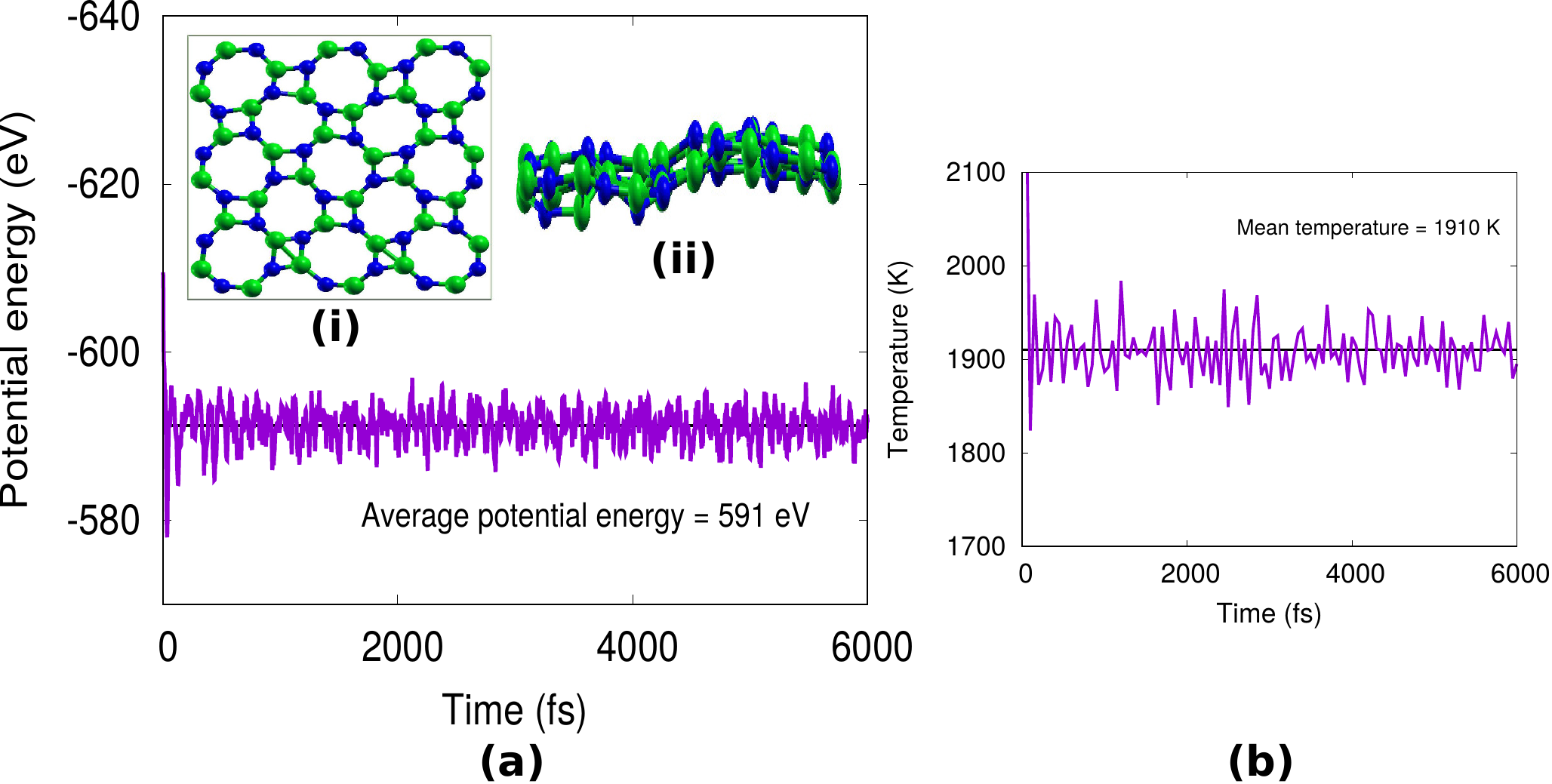}
\caption{Variation of (a)~ionic potential energy and (b)~system temperature as a function of simulation time for 3~$\times$~3~$\times$~1 supercell 
of {\it{o}}-ML of BN, simulated in steps of 1~fs for 6~ps at 1900K.}
\label{MD}
\end{figure}

To investigate thermal stability of {\it{o}}-ML of BN, an ab-initio canonical ensemble MD simulation using Nos$\acute{e}$-Hoover thermostat, 
as implemented in
VASP, is performed. A  3~$\times$~3~$\times$~1 (72 atom) supercell of {\it{o}}-ML of BN is used.
The thermal stability is seen up to 1900K, wherein the temperature of the system is varied in steps of 100K to save the computational efforts.
The {\it{o}}-ML of BN is simulated for 6~ps in steps of 1~fs at 1900K.
The final structure is found to be distorted but the basic skeleton of the structure remains the same as shown in insets (i) and (ii) of Fig.~\ref{MD} (a).
Variation of potential energy (in eV) and mean temperature (in K) of the system with time (in fs) are shown in Figs.~\ref{MD} (a) and (b) respectively.
The average potential energy is -591~eV amounting to 0.50~eV per BN pair higher than the planar structure of {\it{o}}-BN.
On ionic relaxation of the distorted structure, it forms symmetric planar structure. Thus, phonon and ab initio MD calculations indicate that 
{\it{o}}-BN possesses very high thermodynamical stability.
Stacking of {\it{o}}-MLs is possible due to van der Waals bonding.

BNC hybrid structures are potential candidates as intercalation material for Li-ion batteries. We have therefore studied various possible 2D octagonal
structures of BN and C.
Three types of doping \textit{viz.}, substitutional, interstitial and adsorption are possible for monolayer.
The stability of {\it{o}}-ML of BN is investigated for various doping and defect studies and results are discussed in supporting information.
Considering all the types of doping and defects, it is clear that {\it{o}}-ML of BN has very high structural stability and hence it can be 
synthesized experimentally.
Higher structural open space and inhomogeneity compared to {\it{h}}-ML should
promote {\it{o}}-ML of BN to be a potential candidate for hydrogen storage applications, gas sensing and dielectric materials applications.
It is found that doping of C is energetically favorable only as substitutional doping of C$_{8}$ octagonal ring.
Ci {\it et al.} have experimentally investigated {\it h}-BNC.~\cite{Ci} Their results indicate two different band gaps for the hybrid structure corresponding to
{\it h}-BN and graphene and they have concluded that the two phases are thermodynamically immiscible separating in 2D domains. Larger domains are preferred via
decrease in total interfacial domain energy. Experimentally observed structure is kinetically stabilized through non-equilibrium growth process. Theoretical calculations
by Yuge depicted strong preference for neighboring B-N and C-C bonds while B-C and C-N bonds are not favored.~\cite{Yuge} No B-B and N-N bonds are formed.
These experimental and theoretical results are for {\it h}-BNC and our results for {\it o}-BNC are similar.
The preference for substitution of octagonal rings suggests agglomeration tendency. Indeed, such agglomeration of C is experimentally observed.~\cite{Ci}
Covalent bonding in case of {\it{o}}-ML of C and ionic bonding in case of BN also play dominant role in doping preference and will be discussed subsequently.

\begin{figure}[!ht]
\centering
\includegraphics[scale=.52]{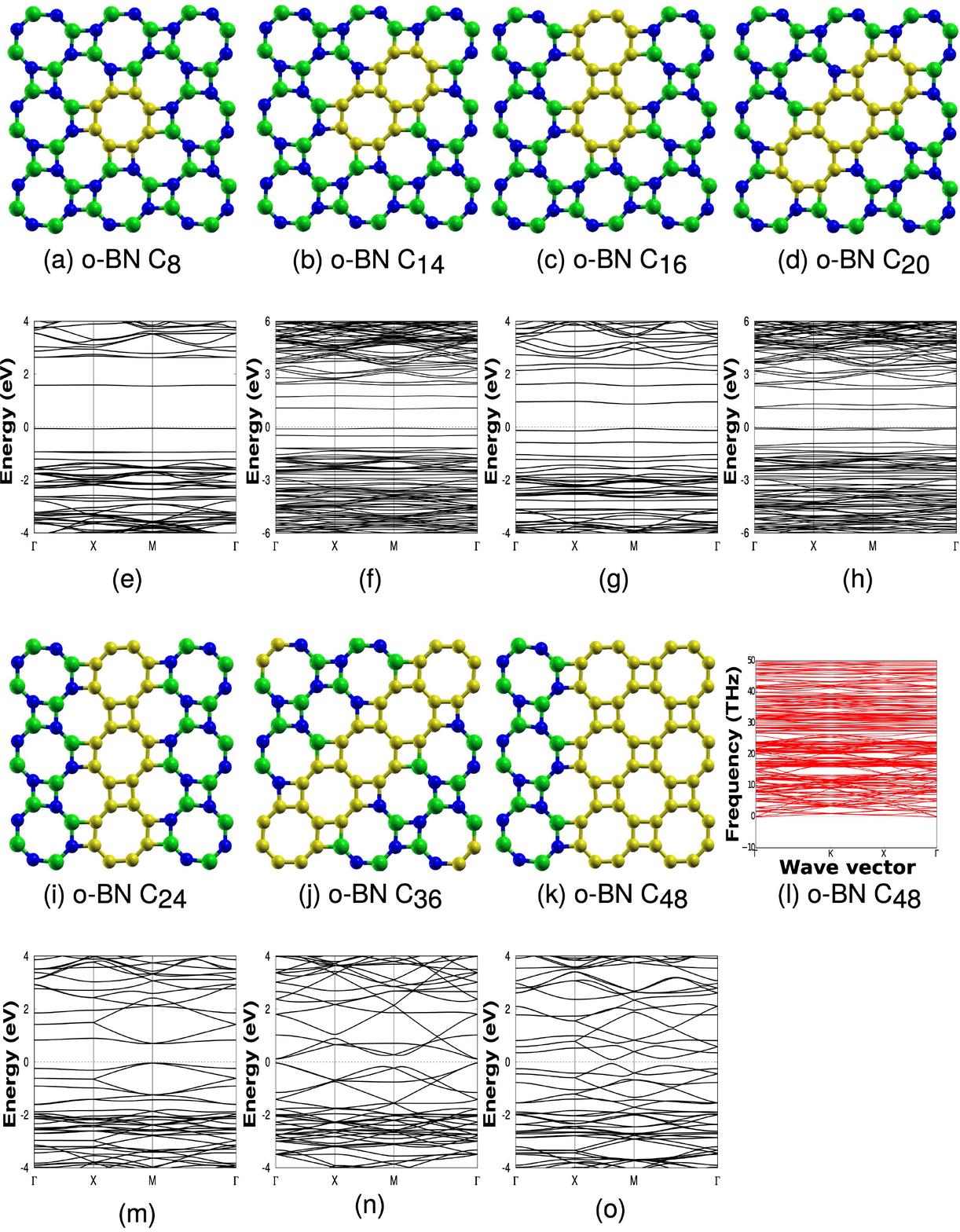}
\caption{Geometries of hybrid assemblies ((a)-(d) and (i)-(k)) and corresponding electronic band structure ((e)-(h) and (m)-(o)) of {\it{o}}-BN 
substitutionally doped with carbon octagonal rings. The green, blue and yellow balls represents boron, nitrogen and carbon atoms respectively. 
(l)~Phonon dispersion plot of hybrid assembly of {\it{o}}-BN-C$_{48}$.}
\label{c-band}
\end{figure}

The lattice parameters and bond lengths of {\it{o}}-MLs of C and BN are close to each other, hence it is expected that the two can accommodate each other easily without leaving strain in patterned hybrid assemblies.
{\it{o}}-ML of BN is substitutionally doped with the units of C$_{8}$ rings.
The substitution of atomic \% of C is varied from 0-100 and minimum energy geometries thus formed are shown in Figs.~\ref{c-band} (a-d) and (i-k).
As expected, the C rings can be easily accommodated.
Dynamic stability of all the hybrid assemblies is confirmed. 
To illustrate, phonon spectra of {\it{o}}-BN-C$_{48}$ hybrid is shown in Fig. ~\ref{c-band}~(l); it has no negative frequencies and conforms the stability of hybrid structure.
Figure~\ref{BE} depicts the variation of cohesive energy (E$^{c}$) and substitutional energy (E$_{sub}$) per C atom on substitution of C rings.
The E$^{c}$ per atom is found to monotonically increase with concentration of C. For pristine BN, the value is 6.78~eV and it increases till 7.48~eV for {\it{o}}-ML of C.
Substitutional energy per C is also found to increase with concentration of C and it is highest for {\it{o}}-ML of C.
This increase in stability can be ascribed to covalent bonding of C-C atoms.
The concentration of C also influences the band gap (E$_{g}$) of the hybrid assemblies (4.13~eV for {\it{o}}-BN to metallic nature for {\it{o}}-C) as shown in Table.~\ref{bader}.
The patterns shown in Fig.~\ref{c-band} indicate that for concentration below 30\% (up to {\it{o}}-BN-C$_{20}$), the C rings are confined and this confinement leads to appearance of discrete {\it{p}} states of C in the forbidden region of {\it{o}}-BN.
For higher percentages, the structures have translational symmetry in one direction.
Of these, {\it{o}}-BNC$_{36}$ shows the least gap because of incomplete BN rings.

\begin{figure}[!ht]
\centering
\includegraphics[scale=.82]{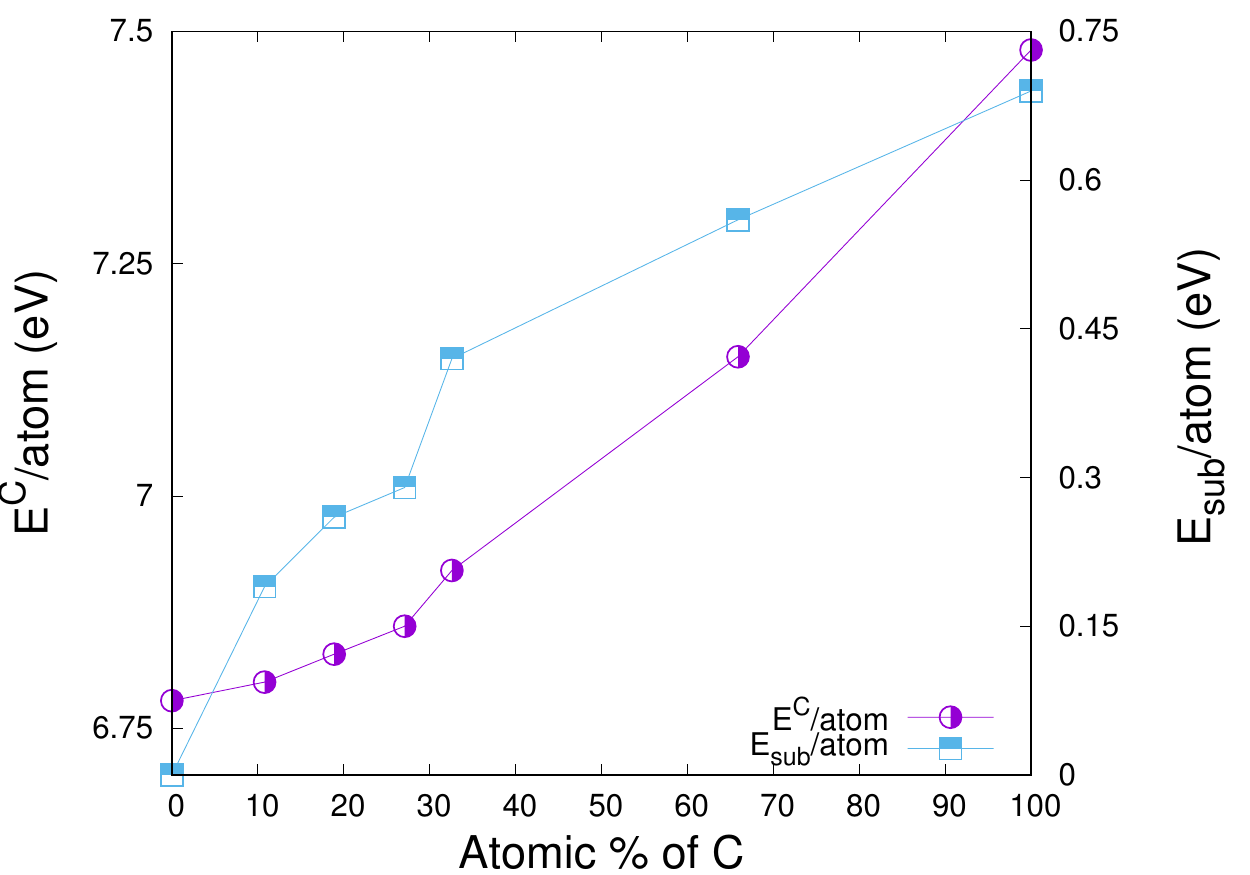}
\caption{Variation of cohesive energy per atom (E$^{c}$ in eV) and substitutional energy per atom (E$_{sub}$ in eV) with the atomic \% of C for {\it{o}}-BNC hybrid assemblies.}
\label{BE}
\end{figure}

The effect of C loading on electronic structure of {\it{o}}-BNC hybrid assemblies is seen in the band structures incorporated in Figs.~\ref{c-band} 
(e) to (h) and (m) to (o).
The deep lying bands around -18~eV (not shown here) are mostly contributed by $p_x$ and $p_y$ states of N.
For pristine BN, valence band (VB) states, from -1~eV to -8~eV, are composed of hybrid {\it{p}}-$p_z$ states of B and N,
whereas conduction band (CB) states are dominated by hybrid states of $p_z$ states of N and {\it{s}} states of B.
For hybrid assemblies, the bands near the Fermi level (in the gap region of {\it o}-BN) are mostly contributed by $\pi$ states of electrons in C rings and this contribution 
increases with C loading.
Larger size of octagonal ring in comparison to hexagonal ring and interaction with interfacial BN states hinders the delocalization of 
$\pi$ states of C and reduces
the band gap to 1.64~eV for {\it{o}}-BN-C$_{8}$ (Fig.~\ref{c-band} (e)).
For the {\it{o}}-ML of BNC hybrids with more than 30\% C substitution ({\it{o}}-BN-C$_{36}$ and {\it{o}}-BN-C$_{48}$), translational symmetry 
along one direction (ribbons) diminishes the band gap.
The band gap of the hybrid structures is seen to depend on the number of B-C and N-C bonds in addition to number of C rings.
The states in the forbidden region are governed by delocalized states of C rings and interfacial B-C or N-C hybridized states.
The charge density plots for hybrid assemblies are shown in supplementary information.

\begin{table}[ht!]
\caption{Band gap, effective average charge on B (Q$^{B}$), N (Q$^{N}$), N atoms connected to carbon (Q$^{N}_{C}$), C (Q$^{C}$), C atom connected 
to B (Q$^{C}_{B}$) and C atom connected to N (Q$^{C}_{N}$), respectively. Charge is expressed in the unit of charge on an electron {\it{e}}.}
\begin{center}
\begin{tabular}{c c c c c c c c}
\hline \hline
 $ System $ & E$_{g}$ & Q$^{B}$ & Q$^{N}$ & Q$^{N}_{C}$ & Q$^{C}$ & Q$^{C}_{B}$ & Q$^{C}_{N}$ \\
\hline
$o$-BN & 4.13 & 3.0 & -3.0 & - & - & - & - \\
C$_{8}$ & 1.64 & 3.0 & -2.98 & -2.82 & -0.09 & 0.64 & -0.82 \\
C$_{14}$ & 1.18 & 3.0 & -2.97 & -2.79 & -0.07 & 0.51 & -0.65 \\
C$_{16}$ & 0.92 & 3.0 & -2.97 & -2.87 & -0.05 & 0.55 & -0.22 \\
C$_{20}$ & 1.19 & 3.0 & -2.95 & -2.75 & -0.07 & 0.54 & -0.68 \\
C$_{24}$ & 1.07 & 3.0 & -2.96 & -2.84 & -0.04 & -0.80 & 0.72 \\
C$_{36}$ & 0.16 & 3.0 & -2.84 & -2.48 & -0.08 & 0.60 & -0.78 \\
C$_{48}$ & 0.60 & 3.0 & -2.91 & -2.82 & -0.02 & -0.73 & 0.52 \\
$o$-C & - & 0.15 & -0.15 & - & - & - & - \\[1ex]
\hline
\end{tabular}
\end{center}
\label{bader}
\end{table}

Substitutional doping of C rings in BN makes C atoms of the ring to adopt some ionic nature which varies with the environment namely the 
arrangement and number of rings.
In all the hybrid and pristine structures, B atoms donate electrons and possess charge Q$^{B}$ equal to +3{\it{e}}.
In case of substitution of C$_{8}$ ring, average charge Q$^{N}$ on N atoms varies from -2.82 to -3.02{\it{e}} with an average value of -2.98{\it{e}}.
The least charge is on N atoms at the interface which are connected to C atoms.
Interestingly, C$_{8}$ ring does not remain neutral because of the presence of B and N atoms. Further the ionicity
increases with higher confinement.
The C atoms connected to B atoms possess 0.64{\it{e}} average charge while those connected to N atoms show -0.82{\it{e}} charge.
Overall average charge is -0.72{\it{e}} per C$_{8}$ ring rather than being neutral in pristine {\it{o}}-C.
The difference in charge transfer among C atoms (each C atoms possess either +0.64{\it{e}} or -0.82{\it{e}}) is due to the fact that,
each interface C atom is triply coordinated with two carbon atoms and the third atom is either B or N of the host.

Substitution of C$_{8}$, C$_{14}$, C$_{16}$ and C$_{20}$ rings are two dimensionally confined systems.
Among them, substitution of C$_{14}$ and C$_{20}$ form diagonally confined hybrid structures while C$_{16}$ is axial substitution as shown in 
Figs.~\ref{c-band} (b), (d) and (c) respectively.
In case of {\it o}-BNC$_{16}$ system, a large variation is seen in charge transfer to/from C atoms connected to N atoms; one of the C atoms has charge 
0.66{\it{e}} while for others it varies from -0.6 to -0.71{\it{e}}.
For this system the charge transfer from C to N atoms is also seen to be least namely -0.22{\it{e}}.
{\it{o}}-BN-C$_{16}$ is the only system that has a complete square of C atoms that too is confined.
When the substitution of C rings is in the form of infinite strips as shown in Figs.~\ref{c-band} (i) and (k), the average charge on C atoms which
are connected to interfacial B and N atoms is reversed in sign, as shown in Table~\ref{bader}.
It is worth mentioning that there is a large variation in the charges on individual C atoms from -0.8 to 0.72{\it{e}}.
In case of {\it o}-BNC$_{48}$ doping, the situation is almost reversed and it can be perceived as single octagonal strip of B$_{4}$N$_{4}$ rings substituted in {\it{o}}-C,
as shown in Fig.~\ref{c-band} (k).
Similar to that of {\it o}-BNC$_{24}$ system, the values of Q$^{C}_{B}$ (-0.73{\it{e}}) and Q$^{C}_{N}$ (0.52{\it{e}}) are reversed in sign.
However, the respective values of Q$^{C}_{B}$ and Q$^{C}_{N}$ are uniform for each atom, as shown in Table~\ref{bader}.
This is compensated by variation in the charge of non interfacial C atoms from 0.52 to -0.73{\it{e}} with most of the C atoms covalently bonded.
For the system with substitution of infinite strip along diagonal, that is in the case of {\it o}-BNC$_{36}$ system, the interface N atoms are least ionic 
and also contain non-complete rings of BN. Hence it is not energetically favored as reflected by cohesive energy per atom.
The charge density plots suggest that no large charge agglomeration happens at the interface of C - BN rings.
Due to higher difference in electronegativity of C and B in comparison to that of C and N, 0.5{\it{e}} of charge is transferred from B to C at 
interfaces and hence B-C bonds at the interface are more ionic in nature than N-C bonds.
Although there is large variation in charge transfer to individual atom depending on the environment, the average charge on each of B, N and C 
atoms does not vary much.

\subsection{Body centered tetragonal structure of {\it{o}}-bulk}\label{bulk1}

\begin{figure}[!t]
\centering
\includegraphics[scale=.52]{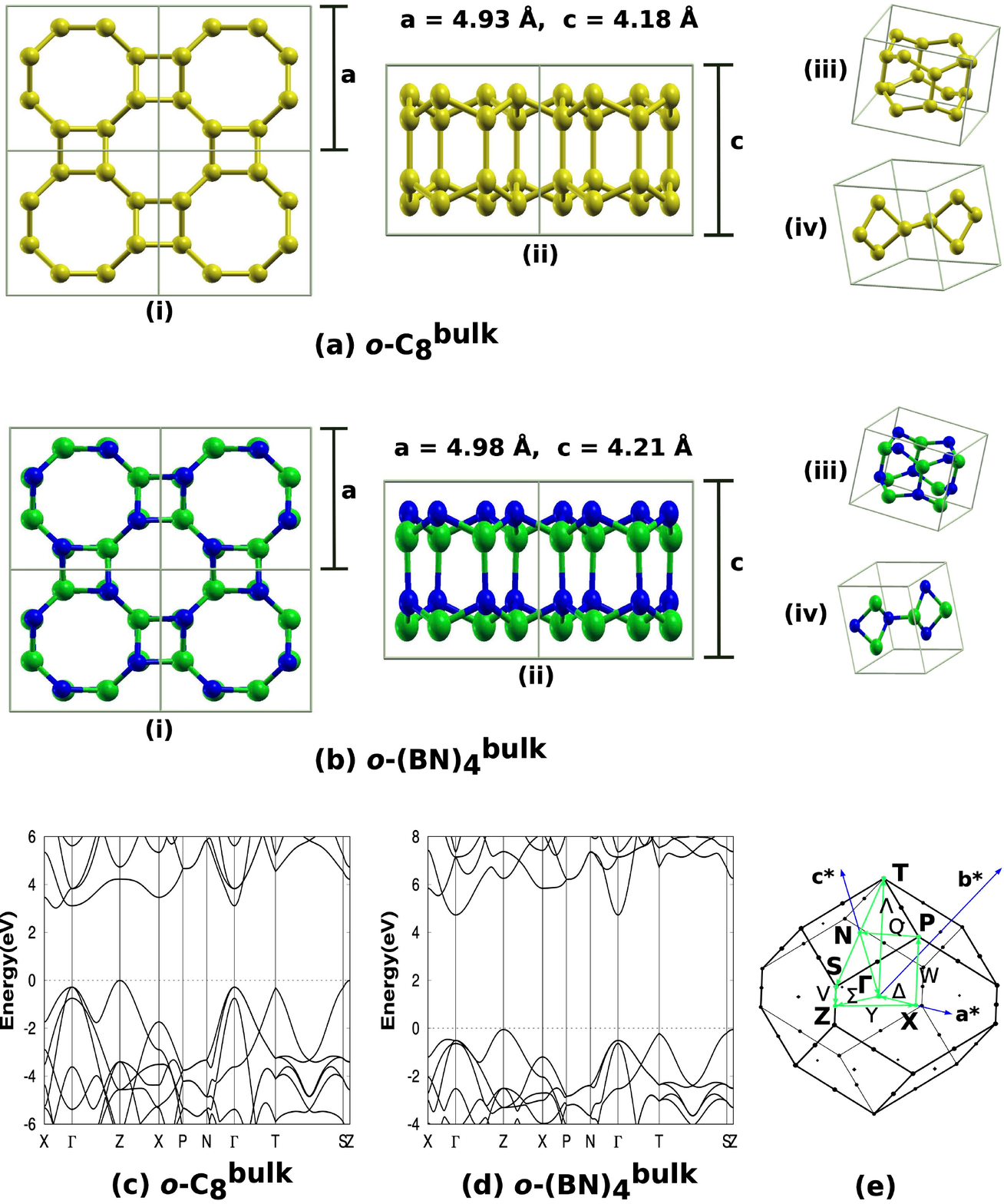}
\caption{Crystal structures of (a)~{\it{o}}-C$_{8}^{bulk}$ and (b)~{\it{o}}-(BN)$_{4}^{bulk}$ in 2$\times$2$\times$2 supercell
are shown in (i)~{\it{xy}}-plane and along (ii)~{\it{z}}-axis respectively. The respective unit cells and primitive cells in 
tetragonal form are shown in (iii) and (iv). The band structure plots of (c)~{\it{o}}-C$_{8}^{bulk}$ and 
(d)~{\it{o}}-(BN)$_{4}^{bulk}$ along symmetry lines (with (e)~high symmetry lines in reciprocal space) show the quasi-direct 
band gap at $\Gamma$ point of 3.12~eV and 4.78~eV respectively. The green, blue and yellow balls represents boron, nitrogen and carbon 
atoms respectively.}
\label{C8}
\end{figure}

As discussed earlier, two types of buckling are possible in the stacked {\it{o}}-MLs and these generate two possible bulk phases of 
bct-carbon; one with buckling of pairs of carbon atoms (bct-C$_{4}$)~\cite{Umemoto, Liu} and other with the zigzag buckling of carbon atoms (bct-C$_{8}$),
henceforth referred to as {\it{o}}-C$_{8}^{bulk}$.
The bct-C$_{8}$ phase is also known as Haeckelite~\cite{Mojica,Brown} bulk structure. Both the phases are from tetragonal family.
A third bulk structure bcc-C$_{8}$ has been reported in the literature~\cite{Liu3,Johnston}.
We have found that the total energy per carbon atom of {\it{o}}-C$_{8}^{bulk}$ is 0.11~eV lower than that of bcc-C$_{8}^{bulk}$ structure.
The body centered tetragonal structures of {\it{o}}-C$_{8}^{bulk}$ and {\it{o}}-(BN)$_{4}^{bulk}$ belong to I4/{\it{mcm}} (140) and 
I4{\it{cm}} (108) group respectively.
The zigzag buckling scheme in the form of octagonal structure (16 atoms) for carbon and boron nitride are shown in Figs.~\ref{C8} (a) 
and (b) while primitive cells (consisting of 8 atoms) are shown in (a)~(iii) and (b)~(iii) respectively.
The bct-C$_{4}$ phase of carbon atom is well documented~\cite{Umemoto, Liu} and has been a matter of interest due to its ability to 
form stable {\it{o}}-ML structure.~\cite{Liu}
On similar lines, {\it{o}}-C$_{8}^{bulk}$ and respective monolayer phases have been reported for various B and Al pnictogen (X$_{8}$Y$_{8}$, 
X = B, Al ; Y = N, P, As, Sb) and GaN.~\cite{Brown,Mojica}
In the present work, we have investigated bct-bulk ({\it{o}}-C$_{8}^{bulk}$) phase of carbon which to the best of our knowledge has not been 
reported in the literature.
Our preliminary studies indicate that all the stable 2D {\it{o}}-ML structures reported in the present work form stable bct-bulk phase.
We would like to point out that the converse is not true in general namely, known stable bct-bulk phase of binary compounds (from III-V 
and II-VI groups) may not result
in stable 2D structures.
Brown et. al and Camacho-Mojica et. al have reported stable bulk-bct phases and respective 2D monolayers of B and Al pnictogen and GaN respectively, but our 
work demonstrates that not all corresponding 2D layers are dynamically stable.~\cite{Brown,Mojica}
We therefore envisage that layered assemblies of {\it{o}}-MLs of materials from IV, III-V and II-VI groups are experimentally realizable.

Table~\ref{bulk} lists the E$^{c}$ per pair and structure parameters of bct-bulk phase of C and BN.
Similar parameters are enlisted for Lonsdaleite and diamond phase of carbon for comparison.
A difference of 0.29~eV/atom between Lonsdaleite and {\it{o}}-C$_{8}^{bulk}$ suggests experimentally plausible synthesis.
The respective bulk phases of C and BN show slightly smaller lattice parameters compared to their 2D {\it{o}}-MLs (Table~\ref{stability}).

\begin{table}[t!]
\caption{Cohesive energy E$^{c}$ per pair (E$^{c}$/XY in eV), lattice parameters (a and c in \AA) and band gap E$_g$ (in eV) 
for various bulk structures.}
\begin{center}
\begin{tabular}{c c c c c}
\hline \hline
System & E$^{c}$/XY (eV) & a (\AA) & c (\AA) & E$_{g}$ (eV) \\
\hline
Diamond & 15.70 & 3.56 & - & 4.11 \\
Lonsdaleite & 15.66 & 2.50 & 4.14 & 3.28 \\
{\it{o}}-C$_{8}^{bulk}$ & 15.09 & 4.06 & 4.06 & 3.12  \\
{\it{o}}-(BN)$_{4}^{bulk}$ & 13.65 & 4.11 & 4.11 & 4.78  \\
{\it{o}}-C$_{8}^{tube}$ & 14.47 & - & 4.21 & 0.00  \\
{\it{o}}-(BN)$_{4}^{tube}$ & 13.17 & - & 4.24 & 1.83  \\
\hline
\end{tabular}
\end{center}
\label{bulk}
\end{table}

\subsection{Nanotubes of {\it{o}}-monolayers}\label{ntube}

\begin{figure}[!ht]
\centering
\includegraphics[scale=.62]{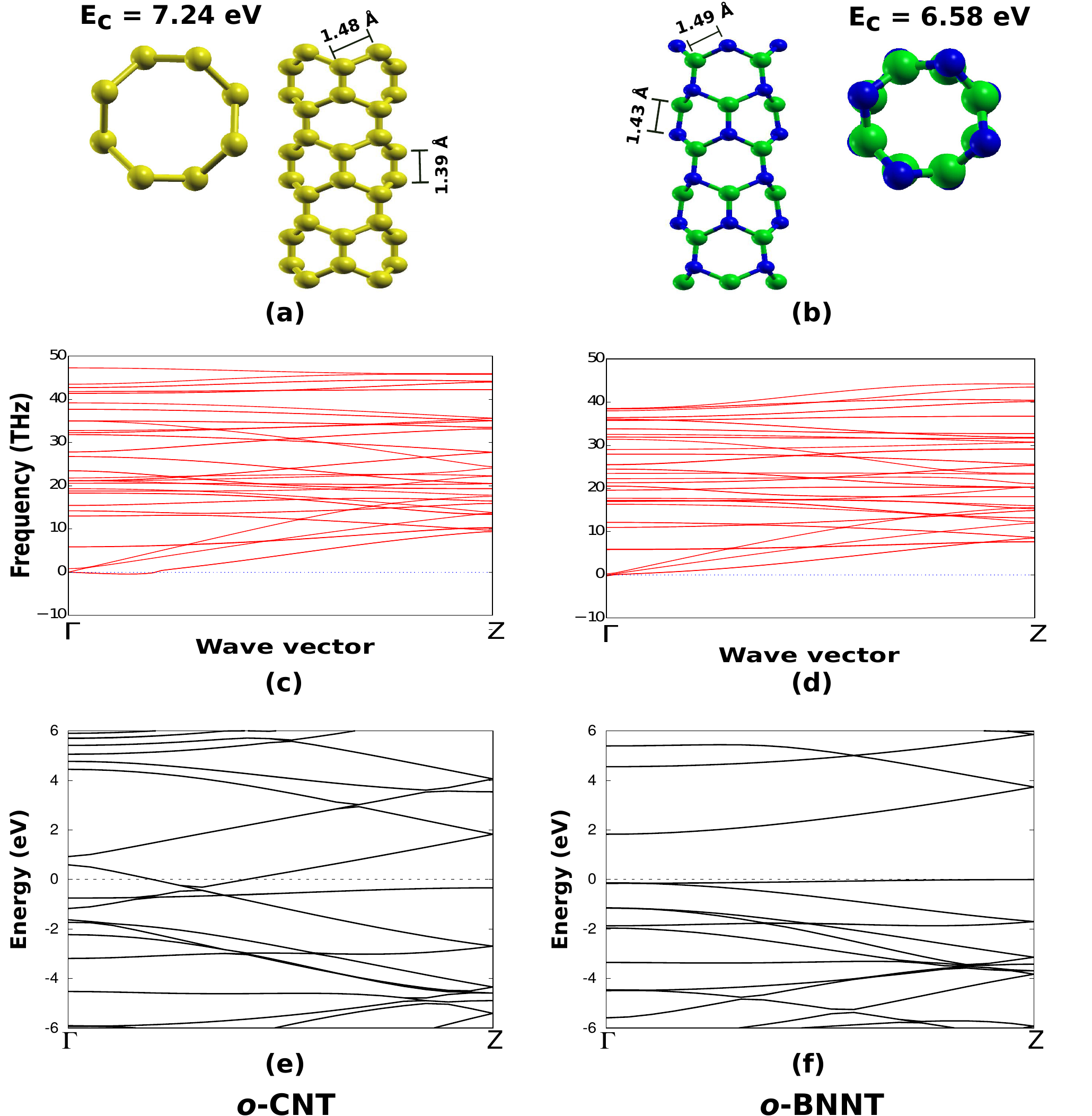}
\caption{Geometry, phonon dispersion spectra and electronic band structure of ((a), (c), (e))~C  and ((b), (d), (f))~BN nanotubes formed by 
stacking {\it{o}}-ML of C and BN respectively.}
\label{t-band}
\end{figure}

Stacking of units of {\it{o}}-MLs and structural buckling along $\textit{z}$-direction can result in formation of nanotubes.
Hence, synthesis of such nanotubes is possible in high pressure compression of {\it{o}}-MLs.
Nanotubes of C ({\it{o}}-CNT) and BN ({\it{o}}-BNNT) thus formed, along with the values of E$_{c}$ are shown in Figs.~\ref{t-band} (a) and (b).
The phonon dispersion spectra of {\it{o}}-CNT and {\it{o}}-BNNT (Figs.~\ref{t-band} (c) and (d)) confirm the dynamical stability of these NTs.
The diameters of {\it{o}}-CNT and {\it{o}}-BNNT are 3.6~\AA\ and 3.57~\AA\ respectively.
It may be noted that the smallest radius reported for stable CNT (obtained by rolling of graphene) is greater than 2~-~4~\AA.~\cite{Qin}
The bond lengths for {\it{o}}-NTs are not identical along zigzag and armchair directions.
The bond lengths of {\it{o}}-CNT and {\it{o}}-BNNT along zigzag direction are 1.48~\AA\ and 1.49~\AA, while along armchair 
direction are 1.39~\AA\ and 1.43~\AA\ respectively. Interestingly, the average bond lengths of {\it{o}}-CNT and {\it{o}}-BNNT 
are 1.43~\AA\ and 1.46~\AA\ respectively, comparable to respective average bond lengths of {\it{h}}-MLs (1.43~\AA\ and 1.45\AA).
If strains of +3.212~\%, -3.212\% and +3.928\%, -1.426~\% are applied for graphene and {\it{h}}-BN lattices and the sheets are 
rolled to form NTs,
bond lengths along armchair and zigzag directions turn out to be comparable to respective {\it{o}}-NTs, as mentioned above.
The calculated strain energies to deform the sheets (difference of total energy of strained and unstrained structure per atom) for 
{\it{h}}-MLs of C and BN are quite small {\it{viz.}}, 0.05~eV/atom and 0.04~eV/atom respectively. However, rolling of sheet to form 
NTs is large as is discussed below.
We have estimated the formation energies for these two processes.
The formation energy is calculated as per atom difference in the energies of respective {\it{o}}-MLs and nanotubes and is a measure 
of energy required for stacking and buckling of {\it{o}}-MLs.
On the other hand, in case of nanotubes formed by rolling {\it{h}}-ML, this formation energy corresponds to rolling of respective 
MLs to form nanotubes.
For BNNT, the calculated formation energies per atom are 0.20~eV and 0.52~eV with respect to {\it{o}}-ML and {\it{h}}-ML of BN respectively.
For CNT these values are 0.24~eV and 0.75~eV with respect to {\it{o}}-ML and {\it{h}}-ML of C respectively.
Comparatively small values of formation energy for {\it{o}}-CNT and {\it{o}}-BNNT suggest strong possibility of experimental synthesis 
of such nanotubes by stacking and compressing respective {\it{o}}-MLs.
Considering the calculated formation energy of rolling of respective strained {\it{h}}-MLs, it is obvious that a large amount 
of energy is to be supplied for rolling of respective {\it{h}}-MLs to form extremely small nanotubes.
The {\it{o}}-NTs reported in this work can be considered as single strip thick zigzag nanotubes of {\it{h}}-MLs of C and BN.
Hence, to obtain NTs of smaller radius, stacking of buckled {\it{o}}-MLs seems to be a plausible way.

Electronic structures of {\it{o}}-CNT and {\it{o}}-BNNT, incorporated in Figs.~\ref{t-band} (e) and (f) respectively,
show that {\it{o}}-BNNT is semiconducting while {\it{o}}-CNT is semi-metallic.
At such extremely small sizes, the band gap of 1.83~eV of {\it{o}}-BNNT is exciting for various applications.
Respective charge density and Bader charge analysis show that bonding in {\it{o}}-BNNT is ionic in nature and is covalent in {\it{o}}-CNT.
The site projected angular momentum decomposed density of states (refer supporting information) and patterns of charge transfer of 
{\it{o}}-CNT and {\it{o}}-BNNT match with the corresponding {\it{o}}-MLs.
These extremely thin NTs are important for fundamental studies due to many factors such as confinement, ionic BNNT versus 
covalent CNT, large electric fields that they can generate, etc.

\subsection{Conclusions}

A class of tetragonally symmetric 2D {\it{o}}-ML family is proposed as the possible stable planar structures after {\it{h}}-ML family.
Among large pool of investigated materials from group II-VI, III-V and IV elements, seven materials (BN, GaN, AlN, C, SiC, GeC, BP) 
(we have previously shown that {\it o}-ZnO is possible~\cite{Prashant})  
have shown possible stable structure
of {\it{o}}-MLs.
Six of these {\it{o}}-MLs belong to P4/{\it{mbm}} (127) and more symmetric {\it{o}}-C belongs to  P4/{\it{mmm}} (123) tetragonal space group.
Energetic, dynamic and thermal stability of these {\it{o}}-MLs along with the electronic structures and bonding are discussed.
We propose that those {\it{o}}-MLs which show soft phonons can stabilize in the form of multilayer assemblies.
Electronic structure of {\it{o}}-MLs show that hybrid {\it{p}}-states dominate near the Fermi level and resulting band gap varies from 4.13~eV to metallic nature.
Most of the {\it{o}}-MLs have direct/quasi-direct band gap in visible/IR/UV region.
{\it{o}}-MLs being comparatively less symmetric with large open volume are potential candidates for dielectric, hydrogen storage and optoelectronic applications.
The doping and/or defect studies suggest substitutional doping of C$_{8}$ rings in {\it{o}}-BN is favorable.
Nearly equal lattice parameters of {\it{o}}-MLs of C and BN favor formation of hybrid assemblies.
The structural stability, energetics and variation in electronic structure of hybrid assemblies are discussed. 

Buckling and stacking of {\it{o}}-MLs leads to stable new bulk phases.
Among few possibilities of buckling, the zigzag buckling between interlayers of {\it{o}}-C leads to bct-C$_{8}$ or {\it{o}}-C$_{8}^{bulk}$ 
phase of carbon (space group I4/{\it{mcm}} (140)) with 3.12~eV indirect band gap.
The high band gap (4.78~eV) {\it{o}}-BN$_{8}^{bulk}$ phase of boron nitride belongs to body centered tetragonal structure  
and I4{\it{cm}} (108) space group.
Similar to BN, other reported binary {\it{o}}-ML materials also show stable {\it{o}}-bulk phase.
The cleavage energy of {\it{o}}-MLs of {\it{o}}-C and {\it{o}}-BN from their respective bulk-phase are within experimentally realizable range.

Stable octagonal nanotubes namely {\it{o}}-CNT and {\it{o}}-BNNT are metallic and semiconducting (band gap 1.83~eV) respectively.
Such NTs can be realized by stacking an octagonal ring on its mirror image or by cutting single column chunk of respective {\it{o}}-bulk.
These NTs are limiting cases of plausible nanotube formation
and can not be formed by rolling MLs.

This work demonstrates the significance of octagonal geometry, possibility of existence of octagonal structures and a need to understand the 
electronic structure of such systems. Moreover, octagonal and other haeckelite structured monolayers can be designed by assembling the geometries of passivated 
clusters obtained from ground state geometries of 
very small clusters.
Additional computational efforts to understand the underlying science and experimental realization of these structures are warranted.

%\begin{suppinfo}
%%Additional results in support of the findings of this work can be found in supporting information.
%Additional information about dynamical stability, thermal properties, defect analysis, charge density and density of states can be found in supporting information.
%\end{suppinfo}

\begin{acknowledgements}
Authors would like to acknowledge DST Nanomission Council, Government of India for financial support through a major research project (DST/NM/NS-15/2011(G)).
\end{acknowledgements}

%\section{Author information}
%
%Dr. Prashant Vijay Gaikwad: (ORCID 0000-0002-0446-9426)
%E-mail: prashant@physics.unipune.ac.in; prashantvg03@gmail.com
%
%Prof. Anjali Kshirsagar: (ORCID  0000-0002-4338-4716)
%E-mail: anjali@physics.unipune.ac.in
%
%Notes: The authors declare no competing financial interest. 
%
%\bibliographystyle{achemso}
%\bibliography{Gaikwad-etal-JPCC}
\bibliographystyle{apsrev4-1} % Tell bibtex which bibliography style to use
\bibliography{Gaikwad-etal} % Tell bibtex which .bib file to use (this one is some example file in TexLive's file tree)

\end{document}